# Meter-Range Wireless Motor Drive for Pipeline Transportation

Wei Liu, *Member*, *IEEE*, K.T. Chau, *Fellow*, *IEEE*, Hui Wang, and Tengbo Yang

Department of Electrical and Electronic Engineering, The University of Hong Kong, Hong Kong, China

This paper proposes and implements a meter-range wireless motor drive (WMD) system for promising applications of underground pipeline transportations or in-pipe robots. To power a pipeline network beneath the earth, both the power grid and the control system are usually required to be deployed deep underground, thus increasing the construction cost, maintenance difficulty and system complexity. The proposed system newly develops a hybrid repeater to enable the desired meter-range wireless power and drive transfer, which can offer a fault-tolerant network with a robust structure for the underground sensor-free WMD while maintaining a high transmission efficiency. Hence, this wireless pipeline network can reduce the maintenance requirement and regulate the flow rate effectively. A full-scale prototype has been built for practical verification, and the system efficiency can reach 88.8% at a long transfer distance of 150 cm. Theoretical analysis, software simulation and hardware experimentation are given to verify the feasibility of proposed meter-range WMD for underground pipeline transportations.

*Index Terms*—Wireless pipeline network, wireless motor drive, meter-range, fault-tolerant network.

## I. Introduction

EVER-INCREASING popularization of electric vehicles (EVs) makes great contributions to green transportations. It can reduce the carbon footprint and promote the carbon neutrality, thus effectively combating the climate crisis. Recently, the rapid development of wireless power transfer (WPT) technology adds the charging safety and flexibility for mobile EVs. Promisingly, the traffic energy internet and the wireless energy trading will kick-off the next research topic in both areas of EVs and WPT [1]. Besides, the WPT technology has been explored aggressively in plenty of significant directions [2] and reached rich outcomes, such as magnetic field focusing [3], [4], multi-receiver WPT [5], [6], and various new wireless schemes. These new schemes included but were not limited to wireless power and drive transfer (WPDT) [7], wireless power-and-move [8], [9], and wireless lighting [10]. In particular, a WPDT technology was reported to directly utilize the drive capability of wireless powers, and it can be readily extended to various applications of wireless power drive, typically such as wireless motor drive (WMD) [7]. Also, this WPDT technology can be borrowed to develop the wireless pipeline transportations and wireless robots [11], [12].

Differing from the wired motor drive [13], capacitive power transfer was deployed to generate the rotor field current for synchronous machines [14], but the WPT serves for the field excitation only without motor drive control. In recent years, the WMD fully utilized the mobility and flexibility of WPT and enabled the wireless motors to work against various harsh environments with good sealing and safety [15]. Nonetheless, this scheme requires three separate receivers for energizing three-phase motor windings, which adds the system volume and complexity. Another WMD scheme was investigated by using a modulation method at the transmitter side and an alternating current (AC) chopper at the receiver side [16], but its voltage stress and power loss of converters shall be concerned. Besides, a more compact self-drive scheme was studied for driving a wireless shaded-pole induction motor [17]. It provides an effective solution for the WMD but fails to avoid the use of active switches at the receiver side.

This paper aims to conceive and develop a wireless pipeline network using the proposed meter-range WMD. It frees the use of sensors, cables or batteries, controllers, active switches, and communication modules beneath the earth, thus reducing the system complexity and maintenance requirements. The use of a hybrid repeater can effectively lengthen the transfer distance for underground WPT. Thanks to the underground deployment, such a hybrid of wireless and wired repeaters will not cause any inconvenience. Importantly, it enables a high-efficiency meter-range WMD and contributes an inherently fault-tolerant operation to withstand various faults of in-pipe receivers and motors. Instead of pulse width modulation [18], the pulse frequency modulation (PFM) strategy [19], [20] is used for the proposed meter-range WMD while reducing both the switching frequency and loss. Finally, mobile EVs can deliver the wireless power across a long range and wirelessly activate the in-pipe motor with speed control. This new wireless pipeline network integrates numerous merits, such as mobility, flexibility, robustness, and maintenance-free.

Section II will introduce the wireless pipeline transportation using the proposed meter-range WMD system. Section III will discuss the system evaluation and the fault-tolerant operation. In Section IV, the simulation and experimentation results will be given to verify the feasibility of proposed meter-range WMD system. A conclusion will be drawn in Section V.

## II. Wireless Power Drive System

### A. Wireless Pipeline Transportation

A schematic of the underground pipeline transportation network is shown in Fig. 1, where a meter-range WMD is recommended for accelerating the in-pipe flow rate. Therefore, a wireless pipeline transportation system will be formed, and its advantages are summarized as follows:

(1) The proposed wireless pipeline transportation system has a robust structure involving no sensors, power grids and batteries, active switches and communication modules beneath the earth.

(2) A fault-tolerant network enables the WMD with high system efficiency at a long transfer distance, and thus the fault-tolerant operation can be inherently realized by limiting the DC inputs at the transmitter side only.

(3) The management of transmitter current can directly regulate the in-pipe receiver current under a specific load.

The energy-carrying EV can locate and park above the major node of the pipeline network for activating the WMD. Generally, the depth of pipeline networks may reach 1~3 m or more [21]. The kilohertz standard for EV wireless charging is hard to deliver the wireless power across such a challenging distance. Considering the requirements of high efficiency, high power capacity and long transfer distance, a hybrid repeater is designed and wins out over other candidates, such as multiple separate repeaters or medium-range WPT systems. Besides, it contributes to a fault-tolerant WPT network for the proposed underground wireless pipeline transportation while causing no inconvenience, thanks to underground deployment.

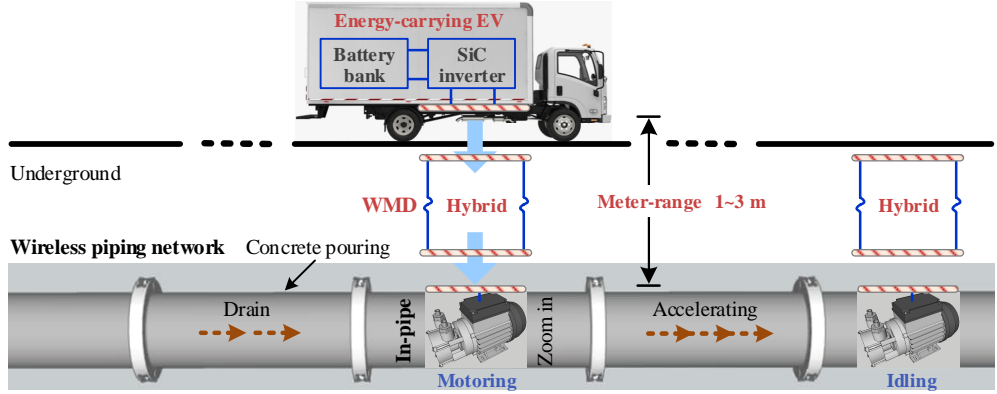

Fig. 1. Schematic of proposed underground pipeline transportation network using meter-range WMD.

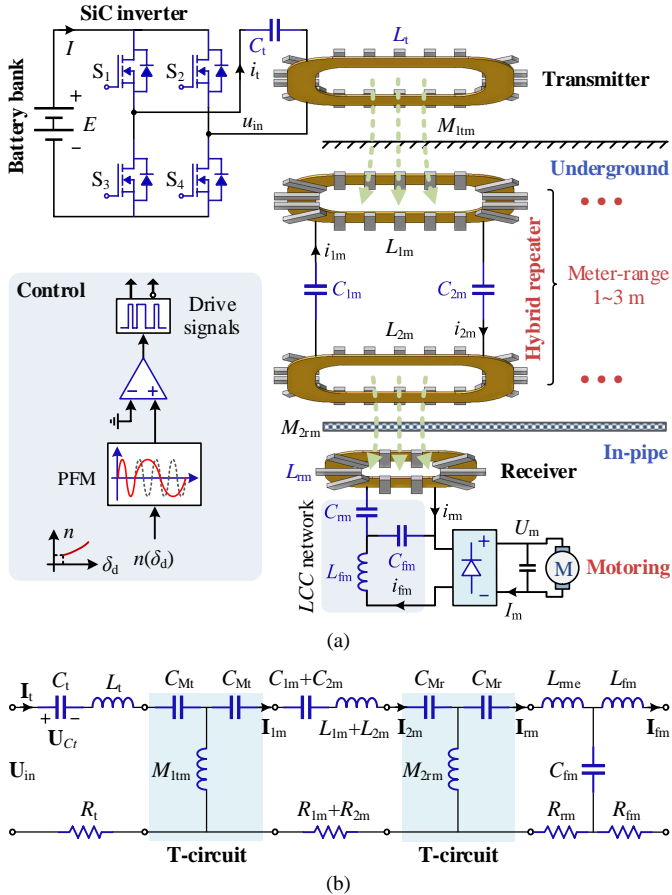

Fig. 2. Proposed meter-range WMD using hybrid repeater for underground pipeline transportation. (a) Topology and control. (b) Equivalent circuit.

### B. Wireless Motor Drive

Fig. 2(a) shows the whole system topology and its PFM control strategy of one set of meter-range WMD. Wherein, the above-ground part is a mobile EV carrying one huge energy storage system, one silicon-carbide (SiC) inverter and one series-compensated transmitter. With the help of WPT technologies, the energy-carrying EV can load the wireless energy to the underground energy-demanding WMD unit. The underground part contains multiple sets of WMD units including the motoring and idling ones, as shown in Fig. 1. Each WMD unit mainly comprises one underground hybrid repeater, one in-pipe receiver, one in-pipe converter and one in-pipe pump. Besides, the in-pipe pump unit integrates one permanent magnet (PM) brushed direct current (DC) motor. The in-pipe receiver adopts an inductor-capacitor-capacitor (*LCC*)-compensation network (including $L_{fm}$, $C_{fm}$ and $C_{rm}$). Since the motor has a variable equivalent load $R_L$ under different motor speeds and loads, such a high-order compensation network aims to provide a load-independent output [10], [19].

In addition, the PFM control strategy can control the output and rotor speed by adjusting its duty ratio $\delta_d$ while suppressing the switching frequency and loss. As for the PFM input voltage, the root-mean-square value $U_{in}$ of the $n_F$th harmonic can be generalized as [19]

$$U_{in} = \frac{2\sqrt{2}E}{\delta_d(2n-1)+(1-\delta_d)(2n+1)} \frac{1}{n_F \pi} \quad (1)$$

where $\delta_d = N_1/(N_1+N_2)$ is the duty ratio, and $N_1$ and $N_2$ are the minimum numbers of modulated pulses at frequencies $f/(2n\pm1)$.

Furthermore, Fig. 2(b) shows the equivalent circuits of the whole WPT network, where two WPT couplers are modeled as two T-circuits. In Fig. 2, four groups of parameters ($R_t$, $L_t$, $C_t$, $i_t$), ($R_{rm}$, $L_{rm}$, $C_{rm}$, $i_{rm}$), ($R_{1m}$, $L_{1m}$, $C_{1m}$, $i_{1m}$) and ($R_{2m}$, $L_{2m}$, $C_{2m}$, $i_{2m}$) represent the coil resistances, resonant inductances, matched capacitances and resonant currents of the transmitter circuit, receiver circuit, and two parts of the hybrid repeater for one motoring unit, respectively. Also, $i_{fm}$ denotes the filter current at the receiver side. Variables $I_t$, $I_{1m}$, $I_{2m}$, $I_{rm}$ and $I_{fm}$ in Fig. 2(b) are the current phasors corresponding to $i_t$, $i_{1m}$, $i_{2m}$, $i_{rm}$ and $i_{fm}$ in Fig. 2(a), respectively. Besides, $M_{1tm}$ and $M_{2rm}$ are the mutual inductances between the transmitter and hybrid repeater as well as between the hybrid repeater and in-pipe receiver, respectively. After the rectification, $C_m$, $U_m$ and $I_m$ are the DC filter capacitor, motor voltage and current, respectively. The idling WMD units have the same definitions of parameters, and the mutual inductance between the transmitter and underground hybrid repeater is $M_{1ti}=0$ due to the long distance. In the WPT network, all resonant parameters should be designed as

$$\omega = \frac{1}{\sqrt{L_t C_t}} = \frac{1}{\sqrt{L_{fm} C_{fm}}} = \frac{1}{\sqrt{\left(L_{rm} - 1/(\omega^2 C_{rm})\right) C_{fm}}} \quad (2)$$

$$\omega = \frac{1}{\sqrt{L_{1m} C_{1m}}} = \frac{1}{\sqrt{L_{2m} C_{2m}}} \quad (3)$$

In Fig. 2(b), $C_{Mtm}=1/(\omega^2 M_{1tm})$ and $C_{Mrm}=1/(\omega^2 M_{2rm})$ are to represent two virtual capacitors in the T-circuits. Also, $L_{rme}$ is the equivalent inductor comprising $L_{rm}$ and $C_{rm}$. Besides, $Z_{CMt}=-Z_{M1tm}$, $Z_{CMr}=-Z_{M2rm}$ and $Z_{Lrme}=Z_{Lfm}=-Z_{Cfm}$. These impedances are represented by referring to their subscripts of components in Fig. 2(b), such as $Z_{CMt}=1/(j\omega C_{Mtm})$ and $Z_{M1tm}=j\omega M_{1tm}$.

The system model can be mathematically generalized as

$$\begin{bmatrix} Z_t & Z_{1t} & 0 & 0 \\ Z_{1t} & Z_{1m} & 0 & 0 \\ 0 & 0 & Z_{2m} & Z_{2r} \\ 0 & 0 & Z_{2r} & Z_{rm} \end{bmatrix} \begin{bmatrix} I_t \\ I_{1m} \\ I_{2m} \\ I_{rm} \end{bmatrix} = \begin{bmatrix} U_{in} \\ 0 \\ 0 \\ 0 \end{bmatrix} \quad (4)$$

where $I_{1m}=I_{2m}$, $Z_{1t}=-j\omega M_{1t}$, and $Z_{2r}=-j\omega M_{2r}$. Also, the self-impedances of transmitter, hybrid repeater (parts 1 and 2) and receiver can be calculated as (i) $Z_t=j\omega L_t+1/(j\omega C_t)+R_t$, (ii) $Z_{1m}=j\omega L_{1m}+1/(j\omega C_{1m})+R_{L12}$, (iii) $Z_{2m}=j\omega L_{2m}+1/(j\omega C_{2m})+R_{2m}$, and (iv) $Z_{rm}=j\omega L_{rme}+1/(j\omega C_{rm})+R_{Lr}$, respectively. Furthermore, $R_{Le}=8R_L/\pi^2$ is the equivalent AC load of motor before the rectification. The loads $R_{Lr}$ and $R_{L12}$ reflected to the in-pipe receiver and hybrid repeater can be respectively derived as

$$R_{Lr} = \frac{(\omega L_{fm})^2}{R_{Le}+R_{fm}} + R_{rm} \quad (5)$$

$$R_{L12} = \frac{(\omega M_{2rm})^2}{R_{Lr}} + R_{1m} + R_{2m} \quad (6)$$

The control of transmitter current can be used to directly regulate the in-pipe receiver current as given by

$$I_t \approx \frac{M_{2rm}}{M_{1tm}} I_{rm} \quad (7)$$

where $I_t$ and $I_{rm}$ are the root-mean-square values of $i_t$ and $i_{rm}$, respectively. It neglects the parasitic resistances in inductors and capacitors.

## III. SYSTEM EVALUATION AND FAULT TOLERANCE

### A. System Characteristic Evaluation

To evaluate the system characteristics, Fig. 3 shows the input impedance characteristics with respect to the operating frequency. With different load resistors, the WPT network has a stable resonant frequency of 85.0 kHz and operates with the zero phase angle (ZPA). Fig. 4 shows the system performances of proposed meter-range WMD system. With a series-/LCC-compensation for the in-pipe receiver, Fig. 4(a) depicts the transmission efficiencies of WPT network, in particular with different filter inductors ($L_{fm}$). Taking account of the varying motor load, the transmission efficiency with an LCC compensation is superior to that with a series compensation. Also, this efficiency when $L_{fm}=0.5L_{rm}$ is much higher than that when $L_{fm}=0.25L_{rm}$. With different mutual inductances, Fig. 4(b) shows the transmission efficiencies with respect to the equivalent motor loads. Because the power under a low load is much larger than that under a high load, the efficiency with mutual inductances ($M_{1tm}$, $M_{2rm}$) is higher at the high-power region.

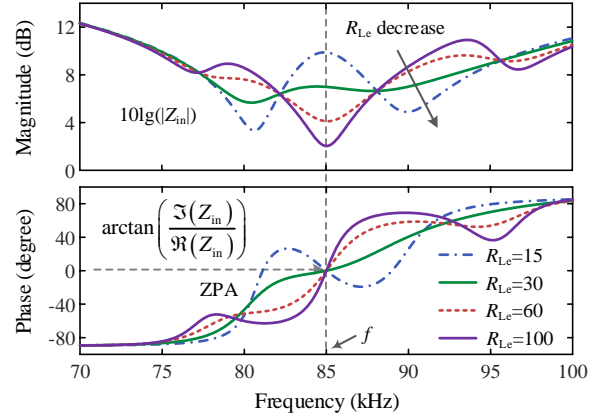

Fig. 3. Input impedance characteristics against operating frequency.

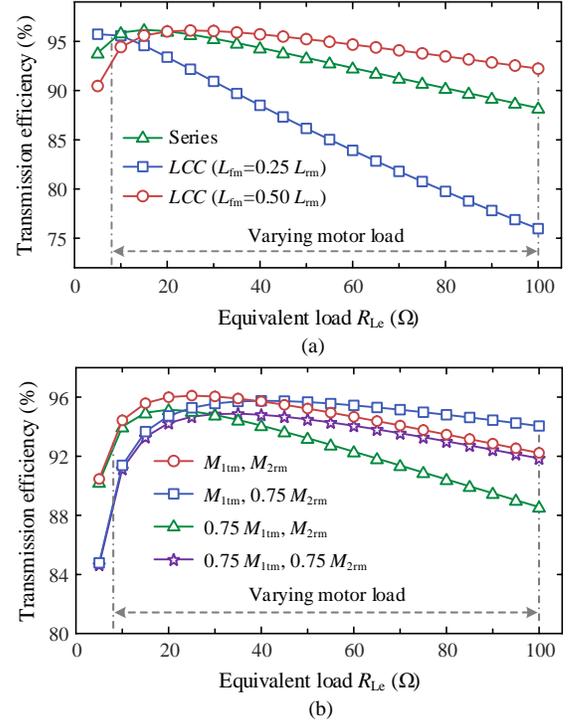

Fig. 4. System performances of proposed meter-range WMD for wireless pipeline transportation. (a) Transmission efficiencies with different compensation networks. (b) Transmission efficiencies with different mutual inductances.

### B. Fault Tolerance Capability

In the proposed meter-range WMD system, all underground parts can be totally sealed for maintenance-free and work in harsh environments. Fig. 5(a) shows the transmitter and receiver currents of proposed meter-range WMD during the normal operation. Various faults, such as the open circuit and the short circuit, may also happen in the in-pipe receivers and motors. Nonetheless, with a simple DC input limiter at the transmitter side only, the fault-tolerant operations can be readily performed. To demonstrate the fault tolerance capability, Fig. 5(b) shows that both the transmitter and receiver currents will be autonomously suppressed to protect the whole WMD system when the fault of motor short circuit or receiver open circuit happens. When the fault of motor open circuit or receiver short circuit occurs, Fig. 5(c) shows that the DC input limiter of the transmitter will prevent the overcurrent in the WMD system.

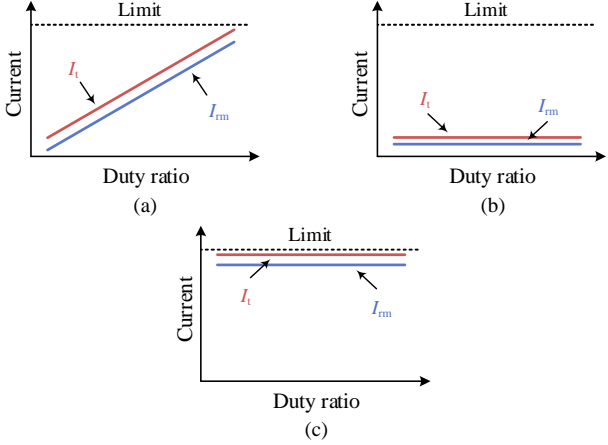

Fig. 5. Fault tolerance capability of proposed meter-range WMD. (a) Normal operation. (b) Motor short circuit or receiver open circuit. (c) Motor open circuit or receiver short circuit (open circuit of filter inductor $L_{fm}$).

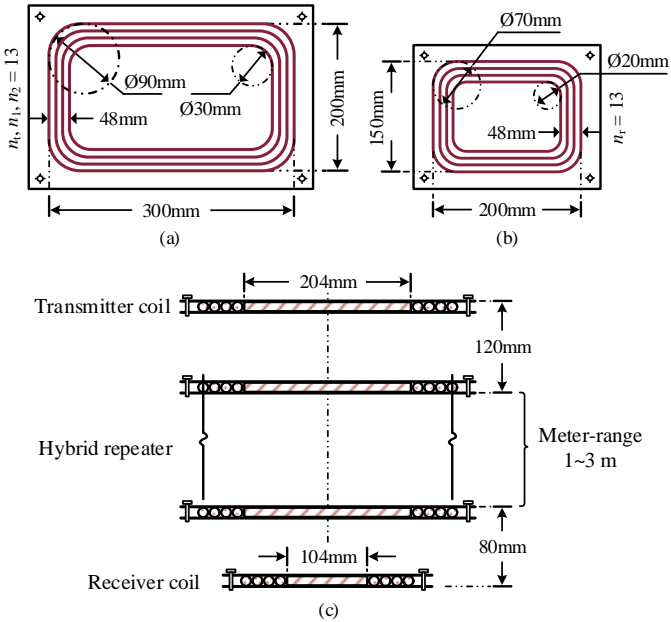

Fig. 6. Geometries of WMD coils. (a) Transmitter and hybrid repeater. (b) Receiver. (c) Displacements.

## IV. RESULTS AND VERIFICATIONS

To verify the feasibility of a wireless pipeline network using proposed meter-range WMD, both the software simulation and hardware experimentation are performed based on a built model and prototype, respectively. Waveforms are measured by an oscilloscope (LeCroy HDO4034A). Powers and efficiencies are measured by using a power analyzer (YOKOGAWA WT3000). The design specifications and parameters of the prototype are listed in Table I. Full scaled geometries of WMD coils are depicted with the detailed dimensions in Fig. 6, where the total distance is 150 cm. For winding the WMD coils, the specification of Litz wire is 600×0.10 mm. Besides, 22 ferrite bars are mounted on the transmitter and hybrid repeater coils, and 20 ferrite bars are used on the receiver coil.

TABLE I
DESIGN SPECIFICATIONS AND PARAMETERS

| Parameter | Symbol | Value |
|---|---|---|
| DC voltage and current limits | $E$, $I$ | 110 V, 7 A |
| Nominal resonant frequency | $f$ | 85.0 kHz |
| WPT coil turns | $n_t$, $n_1$, $n_2$, $n_r$ | 13 |
| Transmitter (Tx) coil inductance | $L_t$ | 86.84 μH |
| Transmitter coil internal resistance | $R_t$ | 0.085 Ω@$f$ |
| Transmitter matched capacitance | $C_t$ | 40.58 nF |
| Hybrid repeater coil inductances | $L_{1m}$, $L_{2m}$ | 86.22, 86.21 μH |
| Hybrid repeater coil internal resistances | $R_{1m}$, $R_{2m}$ | 0.085 Ω@$f$ |
| Hybrid repeater matched capacitances | $C_{1m}$, $C_{2m}$ | 40.68, 40.81 nF |
| Receiver (Rx) coil inductance | $L_{rm}$ | 72.30 μH |
| Receiver coil internal resistance | $R_{rm}$ | 0.08 Ω@$f$ |
| Receiver matched capacitance | $C_{rm}$ | 96.98 nF |
| Receiver filter inductance | $L_{fm}$ | 36.15 μH |
| Receiver filter internal resistance | $R_{fm}$ | 0.02 Ω@$f$ |
| Receiver filter capacitance | $C_{fm}$ | 96.98 nF |
| Transfer distances of Tx and repeater | $D_{1tm}$, $D_{1ti}$ | 12 cm, ∞ |
| Transfer distances of repeater and Rx | $D_{2rm}$, $D_{2ri}$ | 8 cm, 8 cm |
| Mutual inductances of Tx and repeater 1 | $M_{1tm}$, $M_{1ti}$ | 13.56, 0 μH |
| Mutual inductances of Rx and repeater 2 | $M_{2rm}$, $M_{2ri}$ | 21.44, 21.44 μH |
| Total transfer distances | $D_m$, $D_i$ | 150 cm, ∞ |

### A. Simulation Verification

By using the finite element analysis (FEA), an AC current source of transmitter input and an equivalent motor load are used and set as $I_t$=20 A and $R_L$=25 Ω, respectively. Without any displacement or misalignment, Fig. 7(a) shows the magnetic flux densities of two WPT couplers along the vertical plane, which indicates that the wireless power can be successfully delivered to the in-pipe receiver across such a long transfer distance, thus activating the WMD. Furthermore, the 3-D plots of flux densities are shown in Fig. 7(b) along the middle parallel plane. By using ferrite bars, the flux densities between the transmitter and part 1 of the hybrid repeater can reach 1.574 mT in Fig. 7(b) (left), and those between part 2 of the hybrid repeater and in-pipe receiver can reach 1.993 mT in Fig. 7(b) (right). All these FEA characteristics verify that the wireless pipeline network is readily achievable to adjust the flow rate by using the proposed meter-range WMD.

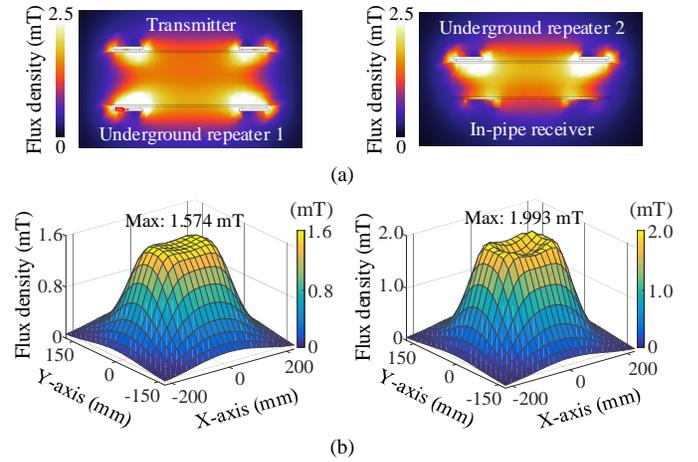

Fig. 7. Magnetic field distributions without displacement or misalignment. (a) Flux densities along vertical plane. (b) Flux densities along middle parallel plane.

Considering the water and soil loss, possible displacement or misalignment may happen in the electrical system, especially for the hybrid repeater underground. Accordingly, the FEA method is also used to study the magnetic field distributions with some displacement and misalignment as shown in Fig. 8 and Fig. 9. On the one hand, with a displacement by 30% coil size of the transmitter, Fig. 8(a) and Fig. 8(b) show the flux densities along the vertical plane and the middle parallel plane, respectively. On the other hand, with a misalignment by 25° slant angle, Fig. 9(a) and Fig. 9(b) also show the flux densities along the vertical plane and the middle parallel line, respectively. All these FEA results well verify that the proposed meter-range WMD using the hybrid repeater underground can still reliably realize the wireless pipeline transportation against possible displacement and misalignment.

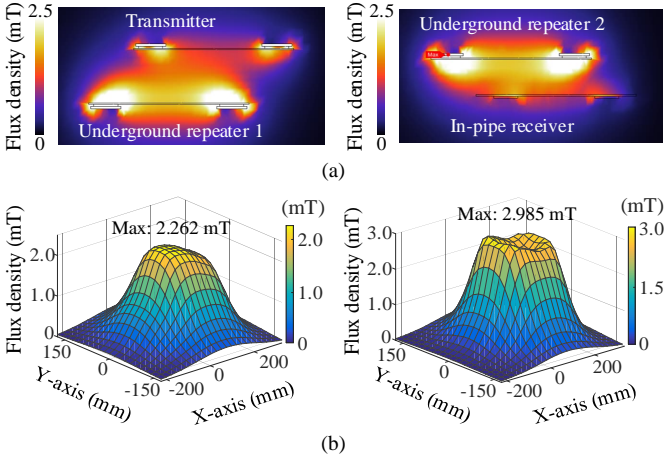

Fig. 8. Magnetic field distributions with displacement due to water and soil loss. (a) Flux densities along vertical plane. (b) Flux densities along middle parallel plane.

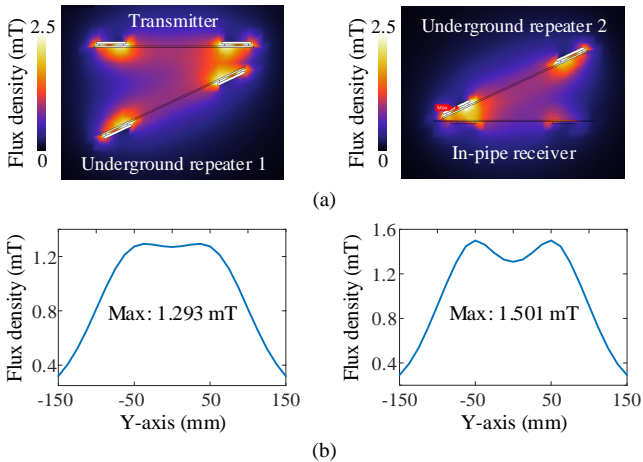

Fig. 9. Magnetic field distributions with misalignment due to water and soil loss. (a) Flux densities along vertical plane. (b) Flux densities along middle parallel line.

## B. Experimental Verification

In Fig. 10, a full-scale prototype was built, and a series of experiments were conducted for verification. As listed in Table 1, the DC voltage and current limits are set as 110 V and 7.0 A, respectively. Fig. 11 shows the measured waveforms of input voltage $u_{in}$, transmitter current $i_t$, repeater current $i_{1m}$, receiver current $i_{rm}$ and motor voltage $U_m$. In Figs. 11(a)–(c), three test conditions are 1633 rpm at no load and $\delta_d=1$, 1245 rpm at the medium load and $\delta_d=4/5$, and 1119 rpm at the rated load and $\delta_d=1$, respectively. By changing the duty ratio $\delta_d$ flexibly, the PFM strategy can modulate two pulse frequencies, typically such as $f$ and $f/3$ [19], and it can effectively control the proposed meter-range WMD system, thus regulating the in-pipe flow rate for underground pipeline transportation as required.

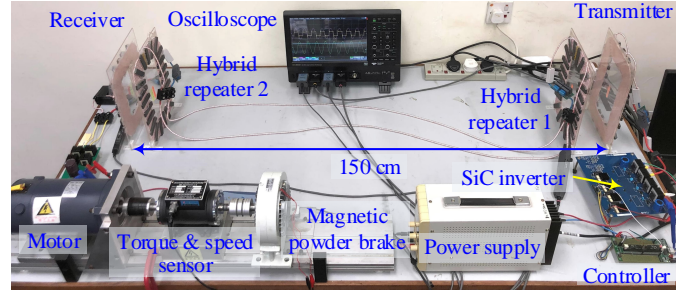

Fig. 10. Experimental setup.

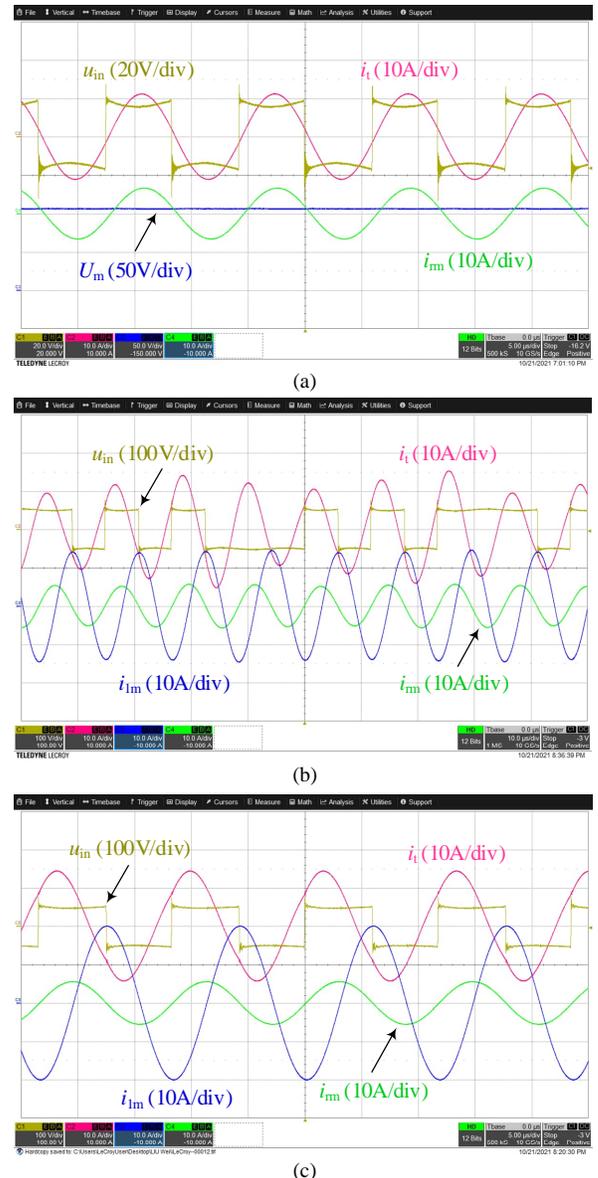

Fig. 11. Measured waveforms of proposed meter-range WMD using the PFM. (a) 1633 rpm (no load). (b) 1245 rpm (medium load). (c) 1119 rpm (rated load).

Furthermore, Fig. 12 shows the measured waveforms of AC inputs of transmitter and DC outputs in the proposed meter-range WMD. Wherein, the DC motor voltage is 87.7 V, and the motor current is 7.2 A. The wireless power can reach up to 631.4 W for driving the in-pipe motor, and the system electrical efficiency is 85.1% under a 1.5-m transfer distance. To access the overall performance, Fig. 13 shows the system electrical efficiency with respect to different power levels. The highest efficiency occurs at 398.6 W, and it can reach up to 88.8%. Also, the system efficiency can always keep above 85.0% from 198.8 W to 638.7 W at a meter-range transfer distance. In the low-power range, a trough on the efficiency is due to the deterioration of zero voltage switching, which is essentially caused by the inductive impedance of equivalent motor load. These measured results well verify the feasibility of proposed meter-range WMD system for wireless pipeline transportations or in-pipe robots.

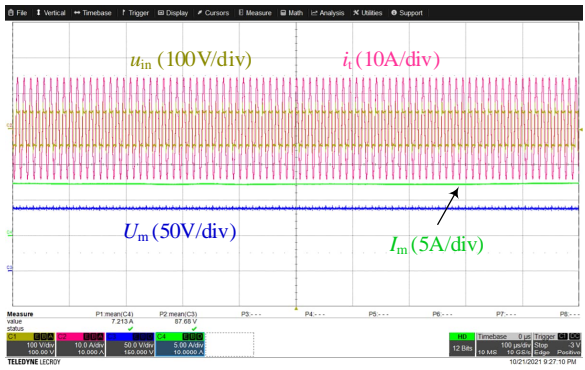

Fig. 12. Measured AC input waveforms and DC output waveforms of proposed meter-range WMD system.

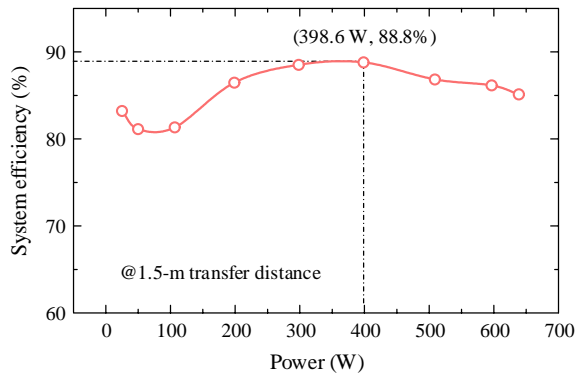

Fig. 13. Measured system efficiency of proposed meter-range WMD system.

## V. CONCLUSION

A high-efficiency meter-range WMD system using a hybrid repeater has been proposed and implemented for promising applications of wireless pipeline networks or in-pipe robots. Thanks to using no sensors, cables and controllers underground, this scheme can effectively reduce the construction cost, maintenance difficulty and system complexity suffered by the cabled pipeline networks. With the help of an underground hybrid repeater, the proposed WMD system enables a meter-range wireless power and drive transfer with inherently fault-tolerant operation. In this meter-range sensor-free WMD, the system electrical efficiency can reach up to 88.8% at a long transfer distance of 150 cm and a power of 398.6 W. Also, it can always maintain over 85.0% from about 200 W to about 640 W. Theoretical analysis, FEA simulation and hardware experimentation are given to verify the feasibility of proposed meter-range WMD for wireless pipeline transportations.

ACKNOWLEDGMENT

This work was supported by a grant (Project no. SFBR 201910159042) from The University of Hong Kong, Hong Kong Special Administrative Region, China.